\documentclass[conference]{IEEEtran}
\usepackage{amsmath,amsfonts}
\usepackage{array}
\usepackage[caption=false,font=normalsize,labelfont=sf,textfont=sf]{subfig}
\usepackage{textcomp}
\usepackage{stfloats}
\usepackage{url}
\usepackage{verbatim}
\usepackage{graphicx}
\usepackage{cite}
\usepackage{tikz}
\usepackage{amssymb}
\usepackage{enumitem}
\usepackage{amsmath}

\def\BibTeX{{\rm B\kern-.05em{\sc i\kern-.025em b}\kern-.08em
    T\kern-.1667em\lower.7ex\hbox{E}\kern-.125emX}}

\makeatletter
\def\ps@IEEEtitlepagestyle{%
	\def\@oddfoot{\mycopyrightnotice}%
	\def\@oddhead{\hbox{}\@IEEEheaderstyle\leftmark\hfil\thepage}\relax
	\def\@evenhead{\@IEEEheaderstyle\thepage\hfil\leftmark\hbox{}}\relax
	\def\@evenfoot{}%
}
\def\mycopyrightnotice{%
	\begin{minipage}{\textwidth}
		\centering \scriptsize
		Copyright~\copyright~2025 IEEE. All rights reserved, including rights for text and data mining, and training of artificial intelligence and similar technologies. Personal use is permitted, but republication/redistribution requires IEEE permission. See https://www.ieee.org/publications/rights/index.html for more information.
	\end{minipage}
}
\makeatother

\begin{document}

\title{KAN-powered large-target detection for automotive radar}
\author {Vinay Kulkarni, V. V. Reddy, Neha Maheshwari \\
\footnotesize \textit {\scriptsize {vinay.kulkarni@iiitb.ac.in, vinod.reddy@iiitb.ac.in, neha.maheshwari@iiitb.ac.in}}\\
\footnotesize \textit{International Institute of Information Technology,
Bengaluru, India} \\
}





\maketitle

\begin{abstract}
This paper presents a novel radar signal detection pipeline focused on detecting large targets such as cars and SUVs. Traditional methods, such as Ordered-Statistic Constant False Alarm Rate (OS-CFAR), commonly used in automotive radar, are designed for point or isotropic target models. These may not adequately capture the Range-Doppler (RD) scattering patterns of larger targets, especially in high-resolution radar systems. Additional modules such as association and tracking are necessary to refine and consolidate the detections over multiple dwells. To address these limitations, we propose a detection technique based on the probability density function (pdf) of RD segments, leveraging the Kolmogorov-Arnold neural network (KAN) to learn the data and generate interpretable symbolic expressions for binary hypotheses. Beside the Monte-Carlo study showing better performance for the proposed KAN expression over OS-CFAR, it is shown to exhibit a probability of detection ($P_{\mathrm{D}}$) of $96\%$ when transfer learned with field data. The false alarm rate ($P_{\mathrm{FA}}$) is comparable with OS-CFAR designed with $P_{\mathrm{FA}}=10^{-6}$. The study also examines how the number of pdf bins in the RD segment affects the performance of KAN-based detection.
\end{abstract}


\section{Introduction}


\IEEEPARstart{W}ith the advent of mmWave and microwave frequencies in automotive radar, the operating wavelength has significantly decreased, making it much smaller than typical object dimensions. This places targets in the optical scattering region~\cite{skolnik1980introduction}, where dimensions exceed 10–20 times the wavelength ($\lambda$). In complex environments, multiple radar reflections are dominated by geometric and specular scattering, influenced by shape, surface, and material properties, posing challenges for automotive radar target detection~\cite{wang2016spatial}.

Consequently, large automotive targets such as cars, SUVs, and trucks fall within this regime with characteristic lengths $>>10\lambda$. Traditionally, the automotive radar signal processing (RSP) pipeline assumes a point or isotropic target model, performing detection on a cell-by-cell basis. However, for large targets, this approach often results in multiple detections per acquisition, including both valid and spurious ones, influenced by the preset false alarm rate. Managing real-time detections is challenging, requiring refinement and association, which is handled by the tracking module after detection.

Target detection is traditionally formulated as a binary hypothesis testing problem, that distinguishes between the null hypothesis (target absence, $H_0$) from the alternate hypothesis (target presence, $H_1$)~\cite{richards2014fundamentals,richards2010principles,skolnik1980introduction}, characterized by the probability of detection ($P_{\mathrm{D}}$) and the probability of false alarm ($P_{\mathrm{FA}}$). A widely used detector in radar signal processing (RSP) is the constant false-alarm rate (CFAR) detector, applied to either $1$-D range data or $2$-D Range-Doppler (RD) map~\cite{richards2014fundamentals,richards2010principles, skolnik1980introduction}. Among various CFAR techniques, ordered-statistics (OS)-CFAR is particularly prevalent in automotive radar~\cite{sun2020mimo}.

Dynamic environment, along with target shape and size, influence radar scattering spread, necessitating an adaptive detection mechanism. This requires accurate detection within minimal dwells, low latency, and self-contained efficiency, reducing reliance on subsequent processing modules~\cite{sun2020mimo}. The need for such adaptability is particularly critical for large targets in automotive scenarios.


CFAR-based adaptive detection techniques, such as truncated statistics CFAR (TS-CFAR)~\cite{tao2015robust}, mitigate statistical outliers in CFAR reference windows. Variance-based, quantile-based (QTS), and QTS augmented with maximum likelihood estimation (QTS-MLE) are among the approaches used for this purpose~\cite{zhou2022robust}. Alternatively, comprehensive CFAR (Comp-CFAR) applies the central limit theorem and log compression for non-coherent accumulation while introducing a protection window for cell-averaging CFAR, improving adaptability and robustness for normally distributed target data~\cite{liu2019research}. Similarly, Greatest of Secondary Detection (GOSD)-CFAR mitigates target masking effects in mmWave radar while maintaining detection probability ($P_{\mathrm{D}}$) without increasing false alarms~\cite{qin2020novel}. However, challenges remain in setting appropriate convolution truncation parameters, adjusting reference windows, and avoiding compression-induced loss of large target spatial extent in automotive radar. Area-Based Combination (ABC) and Area-Based Distribution (ABD) CFAR detection extend CFAR by convolving reference windows (instead of individual reference cells) with fixed kernels to determine detection thresholds~\cite{wei2022area}. While these methods capture target spatial extent, selecting appropriate kernel sizes remains a challenge.


The use of deep learning (DL), convolutional (CNN) and recurrent neural networks (RNN) have proven effective in representation learning by simplifying complex data~\cite{goodfellow2016deep}. Their application in radar systems for target detection is becoming more prevalent~\cite{yang2021novel,wang2019study,stroescu2020combined,brodeski2019deep,chen2023improved}. In~\cite{yang2021novel}, CA-CFAR detections obtained with a reduced threshold (higher detections and false alarms) undergo target classification by a CNN, effectively maintaining increased detection rates while reducing false alarms. Truncated statistics are also augmented with neural networks (TS-NN CFAR)~\cite{dong2024robust} to outperform OS-CFAR and traditional mean-level algorithms. LSTM in~\cite{kulkarni2023detection} improves detection of closely-spaced targets with CFAR, while~\cite{roldan2024deep} uses lidar-based ground truth for extended targets with a Doppler encoder and 3D CNN. Both methods increase computational load and rely on external sensors or manual labeling, making large target detection challenging. 

Studies have explored CNN-based binary detectors with sliding window sizes larger than the target echo in RD map~\cite{wang2019study}. A deep radar detector developed to process the complex IF signal for target detection and localization~\cite{brodeski2019deep}. PointNet-based CFAR detection has been applied to multi-object detection and sea clutter classification via segmentation of raw complex pulse echo data~\cite{chen2024pointnet}. A VGGNet-inspired CNN, combined with region search and probability sorting using Intersection-over-Union (IoU), is used for target identification~\cite{xie2019radar}. In ~\cite{stroescu2020combined}, Faster R-CNN is employed for object detection, along with particle filtering for enhanced tracking. However, these techniques exhibit higher complexity, and face challenges with segmentation labeling. Furthermore, the lack of explainability in neural networks raises regulatory concerns for automotive applications. 

This work tackles large-target detection using Range-Doppler (RD) segment-based detection, unlike conventional cell-by-cell techniques. An RD segment with a detected target offers enriched scattering coverage for a better target representation.

The binary hypothesis problem for RD-segment-based detection is formulated to emphasize the need for a dynamic decision criterion. We use the Kolmogorov-Arnold Network (KAN)~\cite{liu2024kan}, inspired by the Kolmogorov-Arnold representation theorem~
\cite{arnol1957functions}, which provides symbolic decision expressions for interpretability, unlike conventional neural networks. The network, trained on synthetic data, is transfer-learned on limited field data to generate hypothesis expressions. We also present the complete detection pipeline that integrates RD-segment-based symbolic evaluation for target detection. The performance of the proposed method is compared with OS-CFAR using both synthetic and field data.

\section{Problem Statement}
Consider a scenario with $K$ large targets with dimensions significantly larger than the signal wavelength. If the $k$th target is modeled with $I_k$ \textcolor{black}{uncorrelated} scatterers, the received signal is given by
\begin{equation}
y(t,l) \approx \sum_{k=1}^K \sum_{i=1}^{I_k} \alpha_{k_i} s(t-\tau_{k,l}^{(i)}) + \eta(t,l),
\label{eq: radar tgt response}
\end{equation}
where ${\alpha_{k_i}}$ is the response from the $i$th scatterer of the $k$th target, $s(t)$ is the transmitted FMCW signal, $\eta(t, l)$ is the additive noise, and
\begin{equation}
\tau_{k,l}^{(i)} = \frac{2}{c}\big(R_k \pm \Delta{R_{k_i}} + (v_k\pm\delta v_{k_i})(l-1)T_{\mathrm{cri}} \big)
\label{eq: Target scatterer echo delay}
\end{equation}
is the delay for the $i$th scatterer of the $k$th target having range and radial velocity uncertainties of $\Delta R_{k_i}$ and $\delta v_{k_i}$ around the target range, $R_k$ and radial velocity, $v_k$. $T_{\mathrm{cri}}$ is the chirp repetition interval. The received down-converted signal after mixing with the transmitted signal is given by
\begin{align}
\label{eq: radar IF signal}
z(t,l) & \approx \sum_{k=1}^K \sum_{i=1}^{I_k} \alpha_{k_i}e^{\jmath [2\pi \{(f_{R_{k}}\pm \Delta f_{R_{k_i}})t+(f_{D_k} \pm \delta f_{D_{k_i}})(l-1)T_{\mathrm{cri}}\}+ \phi_k]} \nonumber\\
& \quad \quad \quad\quad\quad\quad\quad\quad\quad\quad\quad\quad \quad \quad + \eta(t,l),
\end{align}
where $\phi_k$ is the constant phase term. Range and Doppler spreads $(\Delta f_{R_{k_i}}, \delta f_{D_{k_i}})$ arise due to the response from various parts of the large target. RD map is obtained from the 2D-Fourier transform of $z(t,l)$ as
\begin{equation}
\begin{split}
Z(f_R,f_D) &= \bigg|\sum_{k = 1}^{K} \sum_{i=1}^{I_k} \alpha_{k_i} \mathbb{I}_{f_R}(f_R - f_{R_k} - \Delta f_{R_{k_i}})\\  
& \quad\quad\quad \mathbb{I}_{f_D}(f_D - f_{D_k} - \delta f_{D_{k_i}}) + \eta(f_R,f_D)\bigg|^2,
\label{eq:RD-map}
\end{split}
\end{equation}
where $\mathbb{I}$ is the indicator function that yields nonzero value in the presence of the target at $(f_{R_k} + \Delta f_{R_{k_i}}, f_{D_k} + \delta f_{D_{k_i}}), \forall i\in [1,I_k]$. The target detection is performed on $Z(f_R,f_D)$. 

\begin{figure}[htp]
    \vspace{-1.9cm}
    \centering
    \includegraphics[trim={1.5cm 4.8cm 3cm 8.0cm},width=8.8cm,height=10.0cm,]{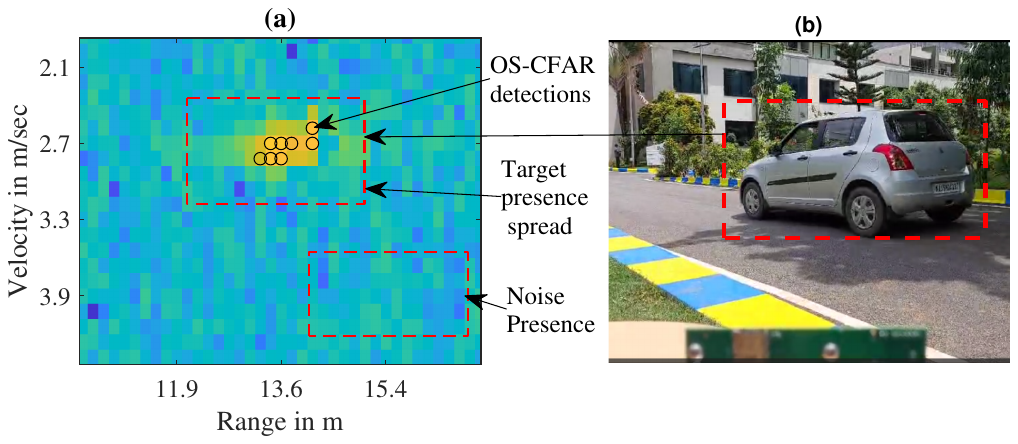}
    \vspace{-4.0cm}
    \caption{(a) RD map with CFAR detections and illustration of RD-segment (b) Corresponding Camera frame.}
    \label{fig:tgt_spread_cfar}     
\end{figure}

Figure~\ref{fig:tgt_spread_cfar} shows the field recording of a $4$m car, with its scattering response in the RD map determined by~\eqref{eq:RD-map}. The presence of $I_k$ scatterers in the target's scattering region leads to a more pronounced spread, especially for large targets. The black circles represent multiple OS-CFAR detections, creating the illusion of multiple targets. 
It is also important to highlight that several cells in the vicinity of CFAR detections are ignored, despite containing vital target information. This effect is pronounced in high-resolution radar, increasing the burden on the association to refine tracks across multiple dwells.

The objective therefore is to accurately capture the full spatial extent of a target, reducing the likelihood of multiple detections for a single target and improving detection efficiency within a single dwell.

\begin{figure}[htp]
    \raggedleft
    \includegraphics[width=8.87cm, height=1.75in]{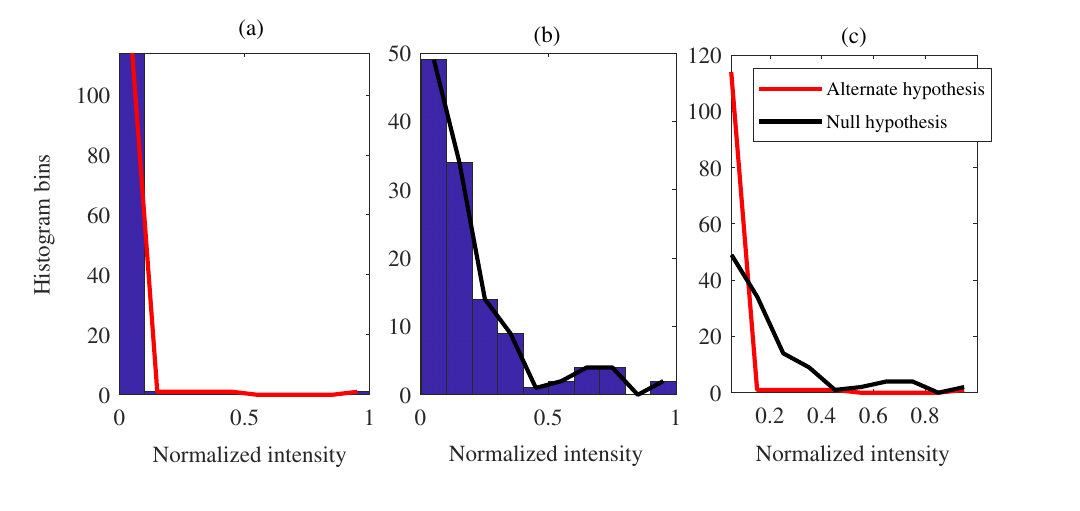}
    \vspace{-1cm}
    \caption{Histogram and its envelope for (a) Alternate hypothesis, (b) Null hypothesis, (c) envelopes for the two hypothesis.}
    \label{fig: Binary hypothesis} 
    \vspace{-0.5cm}
\end{figure}

\section{Proposed Approach}
\subsection{Construction of the Binary hypothesis}
To employ the complete response from large targets in the RD map for detection, we define a localized region called an RD segment which encompasses the typical spatial extent of automotive vehicles, as illustrated by the red rectangle in Fig.~\ref{fig:tgt_spread_cfar}. This segment is designed to cover about $6$m in the range dimension and $2$m$/$s in the Doppler dimension. In contrast to testing individual cells, an entire RD segment is subjected to testing and swept across the RD map to detect all targets within radar range. 

It is well known that the down-converted receive signal $z(t,l)$ follows distinct distributions under each hypothesis:
under $H_0$ (noise-only), it is modeled as $z(t,l) \sim \mathcal{CN}(0,\sigma^2)$ ($\sigma$ is the standard deviation), while under $H_1$ (target present, modeled as Swerling-3), it follows $z(t,l) \sim \chi_4^2 + \mathcal{CN}(0,\sigma^2)$, $\chi_4^2$ denotes Chi-squared distribution of degree $4$. Post square-law detection, the RD cells $Z(f_R,f_D)$ under $H_0$ follow exponential distribution~\cite{Papoulis2002probability,richards2010principles}. However, under $H_1$, the distribution does not converge to a closed-form expression, necessitating a study of its empirical probability density function (pdf). Figure~\ref{fig: Binary hypothesis} presents histograms obtained over several min-max normalized RD segments. In Fig.~\ref{fig: Binary hypothesis}(a), target presence causes most values to concentrate in the first bin, with minimal presence in higher bins. Fig.~\ref{fig: Binary hypothesis}(b) shows a slower decay under the null hypothesis. While both histograms are observed to follow exponential distributions, Fig.~\ref{fig: Binary hypothesis}(c) highlights their differing decay rates $\lambda_0\neq\lambda_1$, with the binary hypothesis as:
\begin{align}
\begin{cases}
H_0: p_z(Z/H_0) &= \prod_{\tilde{n}=0}^{\Tilde{N}-1} \lambda_0 e^{-\lambda_0z_{\tilde{n}}}\\
H_1: p_z(Z/H_1) &= \prod_{\tilde{n}=0}^{\Tilde{N}-1} \lambda_1 e^{-\lambda_1z_{\tilde{n}}},
\end{cases}
\label{eq:Exponential pdf hypothesis}
\end{align}

where $z_{\tilde{n}} \in Z(f_R,f_D)$ are the $\tilde{N}$ cells in the RD segment. In this work, Kolmogorov-Arnold Network (KAN) is employed to learn these distributions for target detection.

\subsection{Kolmogorov–Arnold Network (KAN)}
\label{KANsection}
KAN is a neural network architecture inspired by the Kolmogorov–Arnold representation theorem, which asserts that any multivariate continuous function  $f(\mathbf{x})$ can be expressed as a finite composition of continuous univariate functions~\cite{braun2009constructive}~\cite{liu2024kan}~\cite{liu2024kan2} 
{\small\begin{equation}
    f(\mathbf{x})=f(x_1,x_2,...x_n) = \sum_{q=1} ^{q=2M+1} \Phi_q\bigg(\sum_{r=1}^{r=M} \phi_{q,r}(x_r)\bigg).
    \label{eq:Kolmogorov Arnold Representation}
\end{equation}}
Here $\phi_{q,r}:[0,1] \to \mathbb{R}$ and $\Phi_{q}:\mathbb{R} \to \mathbb{R} , \Phi_q = \left\{\phi_{q,r}\right\}, \forall r =[1, n_{in}]$ \textcolor{black}{and} $q\in[1, n_{out}]$ define the KAN layer mapping from an $n_{in}$-dimensional input to an $n_{out}$-dimensional output~\cite{braun2009constructive}. In~\eqref{eq:Kolmogorov Arnold Representation}, the inner functions form a KAN layer with $n_{in}=M$ and $n_{out}=2M+1$, while the outer functions constitute another KAN layer with $n_{in}=2M+1$ and $n_{out}=1$, resulting in a composition of two KAN layers. Extending this to stack $Q$ KAN layers enables deeper Kolmogorov-Arnold representations, expressed as $f(\mathbf{x}) = (\Phi_Q ...\Phi_1 \Phi_0) \mathbf{x}$. Like Multi-Layer perceptrons (MLPs), KANs have a fully connected architecture, however, they replace fixed activation functions at nodes (``neurons") with learnable $1-$D  activation functions $\phi_{q,r}$ on edges (``weights"). These learnable functions $\phi_{q,r}$ are parameterized by splines~\cite{liu2024kan} and are given by
\begin{align*}
\phi(x) = b(x) + \mathrm{spline}(x) \\
 b(x) = \mathrm{silu}(x) = x/(1+e^{-x}),\\
 \mathrm{spline}(x) = \sum_{i} c_i B_i(x).
\end{align*}
 where $c_i$ are trainable parameters and $B_i(x)$ are B-spline functions~\cite{liu2024kan}, with $i$ controlling the spline order and grid points. Additionally, KAN exploits frequent multiplicative interactions between subnodes and nodes to capture the multiplicative structures in data, a property known as MultKAN~\cite{liu2024kan2}. Consequently, KAN outperforms MLP in accuracy with fewer parameters while providing superior interpretability~\cite{liu2024kan}.

\subsection{Construction of KAN for target detection}
\begin{figure}[htp]
    \raggedleft
    \includegraphics[trim={1cm 6cm 1cm 6.15cm},width=0.48\textwidth, height=3.75cm]{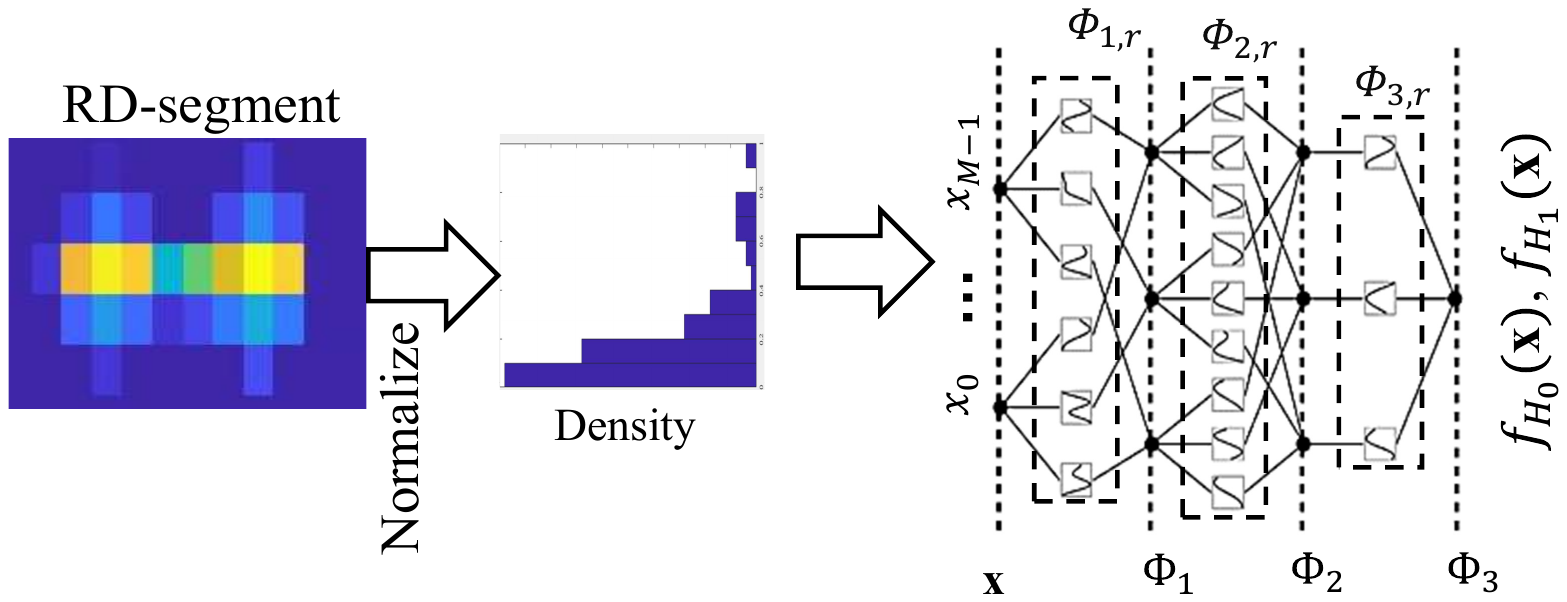}       
    \caption{Illustration of 3-layer KAN with histogram input.}
    \label{fig:KAN_histogram}     
\end{figure}

 In the context of large-target detection, we employ KAN to model the distribution within RD segment. Since KAN constructs univariate functions from each input parameter to the layer's output, feeding the entire RD segment to KAN will require several connections, thus requiring a large amount of training data. Instead, we exploit the contrast in the pdf of RD segments without and with the target spread as observed in~\eqref{eq:Exponential pdf hypothesis}. The histogram of the RD segment with $M-$bins will be normalized and fed to KAN as shown in Fig.~\ref{fig:KAN_histogram}.

This work employs a single-layer of KAN to identify the symbolic expression with $M$ univariate functions that is equivalent to the test statistic for the problem in~\eqref{eq:Exponential pdf hypothesis}. Node and Edge pruning are used during training to discard insignificant univariate functions from inputs. For a test RD segment, the computed histogram bin values are used to evaluate the symbolic expression to determine the presence of the target. The outcome of trained KAN on synthetic and real data is discussed in Sec.~\ref{PerfAnalysis}.

\subsection{Design of RSP detection pipeline}
\begin{figure}[htp]
    \centering       
    \includegraphics[trim={0.6cm 1cm 2.42cm 0.75cm},width=0.465\textwidth]{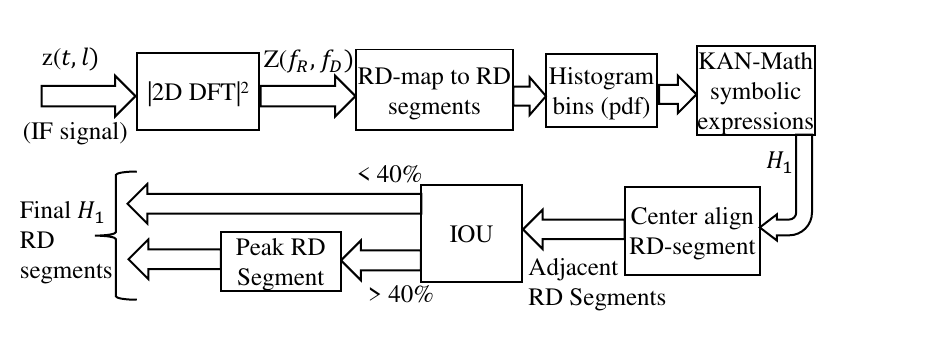}     
    \caption{Proposed detection pipeline.}
    \label{fig:detection_pipeline}     
    \vspace{-0.25cm}
\end{figure}

The detector identifies several neighboring RD segments that cover the large target. We therefore propose the detection pipeline shown in Fig.~\ref{fig:detection_pipeline} that refines the detections with the following considerations:
\begin{itemize}
    \item Once the RD segments are extracted from the RD-map, their $M$ histogram bins serve as inputs to the KAN-driven symbolic expression, facilitating the classification between the null and alternate hypothesis.
    \item RD segments classified as alternate hypothesis are re-aligned to position their center RD-cell at the maximum value. This ensures that the scattering spread of the target, centered around the peak value, is fully captured.
    \item Realigned segments are processed with a $40\%$ Intersection over Union (IoU), selecting the peak RD-segment while retaining lower-IoU segments as separate targets.
\end{itemize}
These steps minimize redundancy in the detected RD segments, thus limiting their number close to the number of targets. As a consequence, the association of these segments across dwells and tracking becomes simpler.

\section{Data Synthesis and Performance Analysis}

\subsection{Field data and Simulated data}
The Medium Range Radar (MRR) specifications in Table~\ref{tab:spec} apply to both simulated and field data. Field data was collected using TI AWR2243 radar with a DCA1000 data capture card. A $4$m-long hatchback car is used for field acquisitions in scenarios that include the vehicle approaching and moving away with and without acceleration. For the specifications in Table~\ref{tab:spec}, the MRR provides a resolution of $0.3516~$m and $0.3044~$m$/$s in range and velocity, respectively. We construct RD segment with $17\times 7$ RD cells that correspond to ($5.98~$m$\times 2.13~$m$/$s).\textcolor{black}{ The Ground truth for real data is derived from manual annotation guided by the target's GPS position.}

\textcolor{black}{Simulated data is generated by synthesizing a Swerling-3 target with $I_k \in [50, 100]$ scatterers, distributed within $R_k \pm \Delta R_k$ (where $\Delta R_k \in [1.5, 3]$ m), with scatterer powers drawn from a $\chi_4^2$ distribution.} RCS values range  from $19$ to $22~\mathrm{dBm}^2$ (side views), $8.7$ to $20.5~\mathrm{dBm}^2$(front views), and $14.4$ to $24.6~\mathrm{dBm}^2$ (rear views)~\cite{kamel2017rcs}, covering cars, SUVs, and trucks at random distances within MRR.

\begin{table}[!ht]
\centering
\vspace*{-2mm}
\caption{FMCW Radar Specification}
\resizebox{\columnwidth}{!}{%
\begin{tabular}{|l|l|l|l|l|l|l|}
    \hline
         $f_0$&FMCW slope&$T_{cri}$&$f_s$&\textcolor{black}{BW}&Samples&Nchirps  \\ \hline
         $77$GHz&$16.67$MHz/$\mu$sec&50$\mu$sec &$10$MHz&\textcolor{black}{$426.68$MHz}&$256$&$128$\\ \hline 
    \end{tabular}
}
\vspace*{-7mm}
\label{tab:spec}
\end{table}

\subsection{Performance Analysis}
\label{PerfAnalysis}
\subsubsection{KAN for hypothesis classification}

Unlike conventional evaluation of detection techniques, large-target detection has to be evaluated within the entire detection pipeline. 
Based on the number of histogram bins drawn from the RD segment, KAN is trained with MultKAN feature with specifications listed in Table~\ref{table: KAN Model details}. KAN provides symbolic expressions at each output node in relation with the input. Detection accuracies presented in Table~\ref{table: Models accuracies} are obtained by evaluating these symbolic expressions for (A) trained and tested on simulated data, (B) trained on simulated data, tested directly on field data, (C) Trained on simulated data, fine-tuned with 14 field samples, and tested on field data. \textcolor{black}{The model handles deviations from~\eqref{eq:RD-map} via transfer learning and can be extended to diverse kinds of targets.}

\begin{table}[!ht]
\vspace{-0.4cm}
    \centering   
    \caption{KAN Model specifications and parameters}
    \begin{tabular}{|l|l|l|l|l|}
    \hline
        Width&Grids&spline&Optimization,Loss&Prune(Node:Edge)\\ \hline
         [$M$,2]&3&3&LBFGS,Crossentropy&0.01:0.03\\ \hline
    \end{tabular}    
    \label{table: KAN Model details} 
    \vspace{-0.3cm}
\end{table}

\begin{table}[!ht]
    \centering
    \vspace*{-3.2mm}    
    \caption{KAN symbolic expression performance}    
    \begin{tabular}{|l|l|l|l|}
    \hline
        Scenario& histogram & train : test(samples)& Accuracy(\%)\\ \hline
        (A)&10 bins&20988:5248 & 99.33\\ \hline
        (A)&5  bins&20988:5248&98.74 \\ \hline
        (B)&10 bins&20988:716&85.89 \\ \hline
        (B)&5  bins&20988:716&84.64 \\ \hline
        (C)&10 bins&20988+14:716&98.04 \\ \hline
        (C)&5  bins&20988+14:716&96.09 \\ \hline 
        \multicolumn{4}{l}{\hspace{-1em}\scriptsize { \textit {(A):  Trained and tested on simulated data.}}} \\
        \multicolumn{4}{l}{\hspace{-1em}\scriptsize { \textit {(B): Trained on simulated data, tested on field data.}}} \\ 
        \multicolumn{4}{l}{\hspace{-1em}\scriptsize { \textit {(C): Trained on simulated data, fine-tuned with field data, tested on field data.}}}
    \end{tabular}
    
    \label{table: Models accuracies}
\end{table}

From Table~\ref{table: Models accuracies}, it is evident that the expressions derived from simulated data show moderate detection performance on field data (Scenario (B)). However, learning the KAN expression with as few as 14 samples in Scenario (C) significantly enhances detection performance, emphasizing the importance of transfer learning. The resulting symbolic expressions for $M=10$ and $M=5$ are obtained as
{\small
\vspace*{-0.2cm}
\begin{equation}
\vspace{-0.5cm}
    -10.288x_0 - 1.14e^{-6}x_1 +7.91 \stackrel{\textstyle >^{H_0}}{\textstyle <_{H_1}}7.5514x_0 - 5.797\\
\label{eq:10bin KAN formula}
\end{equation}

\begin{equation}  
     -2.12\mathrm{e}^{-8}x_1 - 1.65\mathrm{e}^{-8} \stackrel{\textstyle >^{H_0}}{\textstyle <_{H_1}}32.607x_0 - 0.00085x_1 -28.818
\label{eq:5bin KAN formula}
\end{equation}}
\textcolor{black}{Either side of the inequality serve as the test statistics for~\eqref{eq:Exponential pdf hypothesis}}. One of these expressions is integrated with the RSP pipeline shown in Fig.~\ref{fig:detection_pipeline}. Notably, the KAN expression relies on only the first histogram bin $x_0$ for its decision in both expressions.

\begin{figure}[htp]
    \raggedright
    \includegraphics[trim={0.65cm 1cm 2.25cm 1cm},width=8cm,height=4cm,]{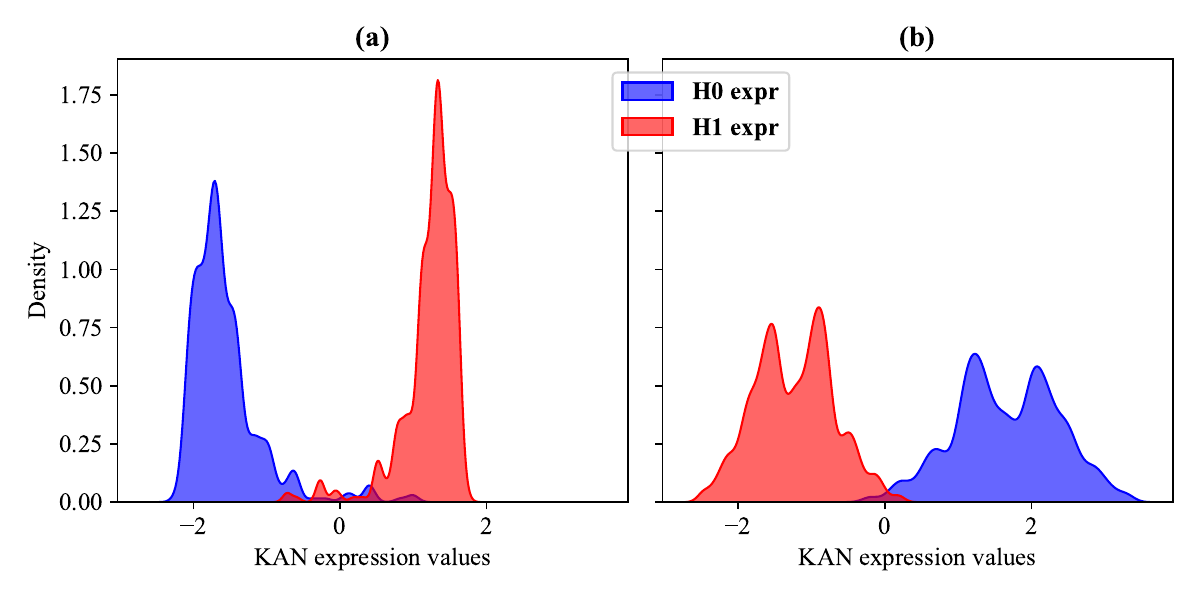}    
    \caption{KDE plot showing KAN separation using $10$ histogram bins\\
    (a) Target RD-segments under test, (b) Noise-only RD-segments under test.}
    \label{fig:KAN_sep_10}
\end{figure}

\begin{figure}[!htp]    
    \raggedright
    \includegraphics[trim={0.65cm 0.9cm 2.25cm 1cm},width=8cm,height=4cm,]{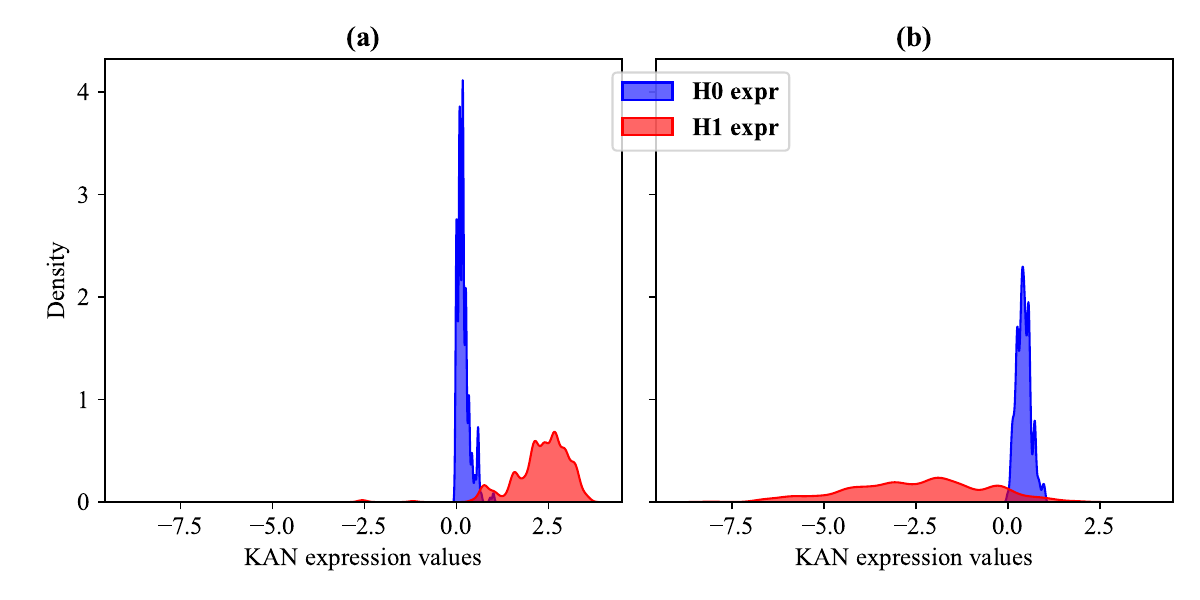}    
    \caption{KDE plot showing KAN separation using $5$ histogram bins\\
    (a)Target RD-segments under test,(b) Noise-only RD-segments under test.}
    \label{fig:KAN_sep_5}
    \vspace{-0.5cm}
\end{figure}

To evaluate the performance of the expressions in~\eqref{eq:10bin KAN formula} and~\eqref{eq:5bin KAN formula}, RD segments reserved for testing are analyzed. A kernel density estimation (KDE) plot is used to visualize the distributions of the evaluated values obtained from the $H_1$ and $H_0$ expressions over multiple samples. Fig.~\ref{fig:KAN_sep_10}(a) presents the KDE plot for target RD segments with $M=10$ bins (using\eqref{eq:10bin KAN formula}), while Figure~\ref{fig:KAN_sep_10}(b) shows the KDE for RD segments without the target. Similarly, KDE plots for the $M=5$ bin (using~\eqref{eq:5bin KAN formula}) KAN are shown in Figures~\ref{fig:KAN_sep_5}(a) and (b).


The plots demonstrate that the expressions effectively separate target and noise-only RD segments, enabling successful detection. For RD segments containing a target, the $M=10$ bin case using~\eqref{eq:10bin KAN formula} yields evaluated values from the $H_1$ expression that form a distribution positioned to the right of the distribution evaluated from the $H_0$ expression as shown in Fig.~\ref{fig:KAN_sep_10}(a). The opposite trend is observed for noise-only RD segments in Fig.~\ref{fig:KAN_sep_10}(b). A similar pattern is seen for the $M=5$ bin KAN case; however, the distribution of evaluated values from $H_1$ expression for noise-only RD segments exhibits greater spread. Additionally, the overlap between the evaluated values of the expressions for $H_0$ and $H_1$ in noise-only RD segments, as shown in Fig.~\ref{fig:KAN_sep_10}(b) compared to Fig.~\ref{fig:KAN_sep_5}(b), suggests a lower probability of false alarm ($P_{\mathrm{FA}}$) for the $M=10$ bin case relative to the $M=5$ bin case.

\begin{figure}[htp]
\vspace{-0.5cm}
    \centering
    \includegraphics[trim={0.6cm 0 0 0},clip,width=9.5cm,height=5cm]{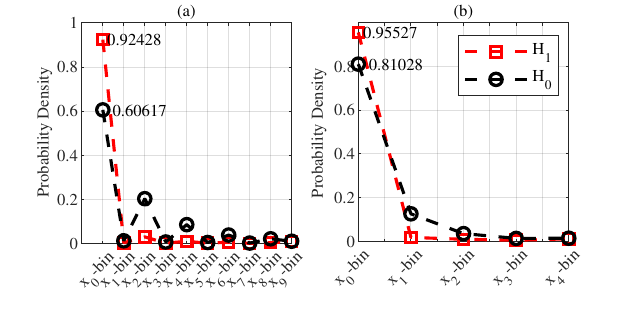}
    \vspace{-1.15cm}
    \caption{Pdf distribution with (a) 10 histogram bins and (b) 5 histogram bins.}
    \label{fig:KAN-histogram_10bins}
    \vspace{-0.18cm}
\end{figure}

To interpret the use of only the $x_0$ bin in~\eqref{eq:10bin KAN formula} and~\eqref{eq:5bin KAN formula}, we analyze target and noise only RD segments with $M=10$ and $M=5$ bins after min-max normalization. The corresponding pdf of RD segments are shown in Fig.~\ref{fig:KAN-histogram_10bins}(a) and Fig.~\ref{fig:KAN-histogram_10bins}(b), respectively. 
In both the subplots, the probability of $x_0$ bin for RD segment with target is $0.92$ or more. This large probability is due to the dominant RD cell normalizing most other cells to $x_0$ bin. Since the pdf is modeled as an exponential distribution for both hypothesis, their decay rates have to be distinct from one another. For $M=10$ bin (Fig.~\ref{fig:KAN-histogram_10bins}(a)) histograms, the decay rates for $H_0$ and $H_1$ using~\eqref{eq:Exponential pdf hypothesis} can be estimated with the probability of the first bin as
\vspace{-0.3cm}
{\small\begin{align}
    H_1: \int_0^{0.1} \lambda_1 \exp(-\lambda_1 t) dt &= 0.9243, & \lambda_1 &= 25.809 \label{eq:lamda1},\\
    H_0: \int_0^{0.1} \lambda_0 \exp(-\lambda_0 t) dt &= 0.6062, & \lambda_0 &= 9.32. \label{eq:lamda0}
\end{align}}

For $M=5$ bin case (Fig.\ref{fig:KAN-histogram_10bins}(b)), the decay rates are obtained to be $\lambda_1=15.538$ and $\lambda_0=8.312$. While KAN does not explicitly provide these decay rates, it learns the distribution by emphasizing the $x_0$ bin of the input pdf and transforming it to maintain the binary hypothesis test inequality. Interpreting the density under $x_0$ for $H_0$ and $H_1$ as an exponential distribution with decay rates $\lambda_0$ and $\lambda_1$ (\eqref{eq:lamda0}, \eqref{eq:lamda1}) offers an alternative approach for target detection.

In one dwell, with approximately $NL$ RD segments or CFAR windows, the real-time computation for the proposed KAN-powered detector is $O(NL\Tilde{N})$ (where $\Tilde{N}$ is the number of cells in an RD segment, assuming $M$ histogram bins with $M < \Tilde{N}$). In contrast, OS-CFAR has a complexity of $O(NL N_{ref} \log N_{ref})$ (assuming quicksort), where $N_{ref}$ is the reference window size. Considering $N_{ref} \approx \Tilde{N}$ highlights a notable computational advantage for the KAN detector. 


\subsubsection{Performance of detection pipeline}
\begin{figure}[ht]
\vspace{-.5cm}
    \raggedleft
    \includegraphics[width=8.85cm,height=9.5cm]{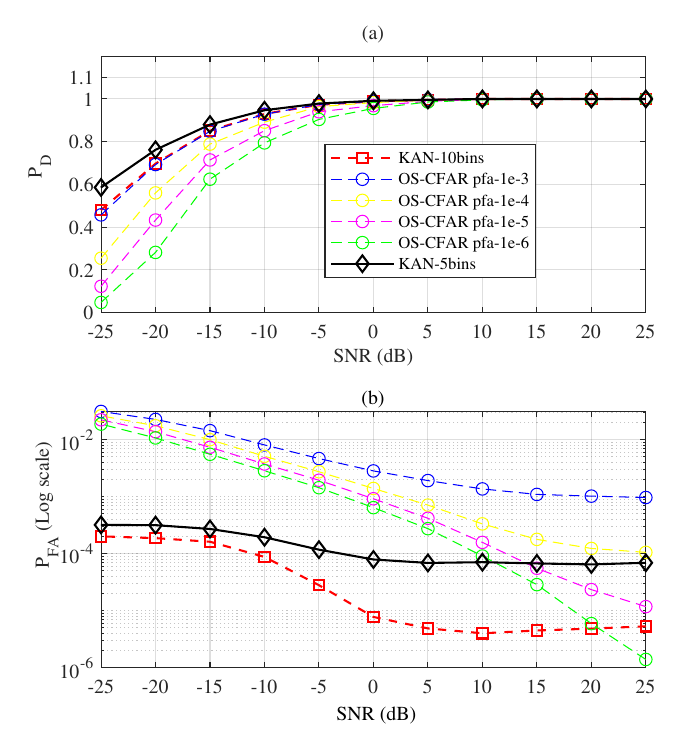}
    \vspace{-1cm}
    \caption{Performance comparison between OS-CFAR and KAN against SNR, (a) Probability of detection, (b) False alarm-rate.}
    \label{fig:KAN_CFAR_pd}
    \vspace{-.5cm}
\end{figure}

\textcolor{black}{For the proposed method, $P_D$ is the percentage of target scattering response RD cells within the ground truth region. In OS-CFAR, a target is detected if at least one CUT within the ground truth region is identified. False alarms ($P_{\mathrm{FA}}$) are detected RD segments or CUTs with no corresponding target response.}
The proposed KAN-powered technique is compared with OS-CFAR with reference window matching the RD-segment and threshold design based on  $P_{\mathrm{FA}} \in [10^{-3}, 10^{-6}]$ using both simulated and field data. \textcolor{black}{For a fair comparison with OS-CFAR, the reference cells used for the threshold computation will be from the same RD segment excluding the CUT and the guard cells $[2,1]$ along range and Doppler axes, respectively.} The $P_{\mathrm{D}}$ and $P_{\mathrm{FA}}$ of these techniques are evaluated over $350$ Monte-Carlo trials in Fig.~\ref{fig:KAN_CFAR_pd} against SNR $\in[-25,25]~$dB.  

As shown in Fig.~\ref{fig:KAN_CFAR_pd}(a), the proposed KAN detector with $M=10$ and $M=5$ bins achieves higher $P_{\mathrm{D}}$ than OS-CFAR detectors designed with $P_{\mathrm{FA}}\in[10^{-3},10^{-6}]$. The $P_{\mathrm{FA}}$ of the proposed technique is observed in Fig.~\ref{fig:KAN_CFAR_pd}(b) to outperform OS-CFAR  particularly for low SNR conditions. The significant improvement in $P_{\mathrm{D}}$ is attributed to the utilization of the full target response, unlike OS-CFAR, which relies on a cell-based approach. Additionally, integrating IoU and centering mechanisms helps further reduce the false alarm rate.

\begin{figure}[htp]
    \raggedleft
     \includegraphics[width=8.5cm,height=7.25cm]{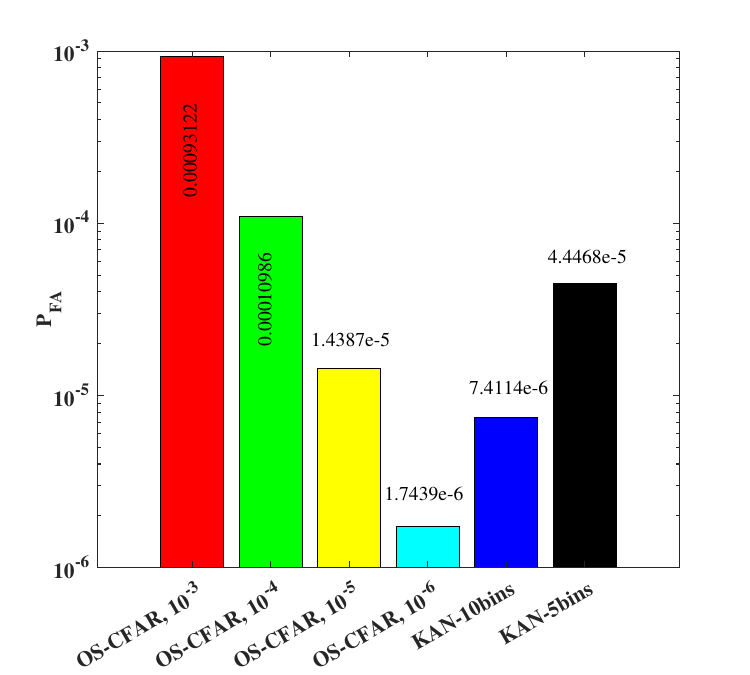}
    \vspace{-0.2cm}
    \caption{\textcolor{black}{$P_{\mathrm{FA}}$ performance of KAN-10, 5 histogram bins vs OS-CFAR designed with $P_{\mathrm{FA}}$ between $10^{-3}$ to $10^{-6}$.}}
    \label{fig:KAN_CFAR_PFA}
    \vspace{-0.6cm}
\end{figure}

On field data, all the techniques detected the large targets present in the scene. We therefore compare only the number of false alarms generated by these techniques. Fig.~\ref{fig:KAN_CFAR_PFA} shows the $P_{\mathrm{FA}}$ comparison between OS-CFAR and the proposed KAN detector with $10$ and $5$ bins. 
OS-CFAR provides false-alarm rates comparable to the values that they were designed for. The $10-$bin and $5-$bin KAN detectors exhibit false alarm rates comparable to OS-CFAR designed with $10^{-6}$ and $10^{-5}$, respectively. 
Combining observations from simulated and field data, the KAN-based pipeline outperforms OS-CFAR, achieving detection rates equal to or exceeding OS-CFAR designed with $P_{\mathrm{FA}} = 10^{-3}$, while maintaining a low false alarm rate of $10^{-5}$ to $10^{-6}$. This performance is achieved with single dwell and  without support of association and tracking, making it highly suitable for automotive target detection. 




\section{Conclusion}
A new radar detection pipeline is presented for large automotive targets, performing detection on RD map segments rather than individual cells to account for target scattering spread. KAN is utilized to learn the symbolic expression, enabling real-time evaluation due to its interpretability. The proposed pipeline is evaluated on both simulated and field data, demonstrating a higher probability of detection and a comparable false alarm rate relative to OS-CFAR.

\bibliographystyle{IEEEtran}
\bibliography{references.bib} 

\end{document}